\documentclass[a4paper,11pt]{article}
\pdfoutput=1      
\usepackage[DIV12]{typearea}
\usepackage[latin1]{inputenc}
\usepackage[english]{babel}
\usepackage{cite}
\usepackage{amsfonts}
\usepackage{amsmath}    
\usepackage{amssymb}
\usepackage{cite}
\usepackage{mathrsfs}
\usepackage{siunitx}
\usepackage{pifont}
\usepackage{units}
\usepackage[normalem]{ulem}
\usepackage{graphicx}
\usepackage{booktabs}
\usepackage[pdftex,colorlinks,pdfpagelabels]{hyperref}
\usepackage[figure,table]{hypcap} 

\hypersetup{
   bookmarksnumbered,
   pdfstartview={FitV},
   pdfpagemode={UseOutlines},
   pdfauthor={Rolf Kappl and Annika Reinert},
   pdftitle={},
   pdfsubject={scientific publication}
   ,citecolor={blue},
   linkcolor={blue},
   urlcolor={blue},
   filecolor={blue}
} 

\DeclareSIUnit[number-unit-product={}]\kpc{kpc}
\DeclareSIUnit[number-unit-product={}]\MV{MV}
\newcommand{\del}{\partial}
\newcommand{\dd}{\text{d}}
\newcommand{\GeVn}{\unit{GeV/n}}
\newcommand{\e}{\text{e}}

\newcommand{\eVdist}{\kern-0.06em}

\newcommand{\GeV}{\text{Ge\eVdist V}}

\allowdisplaybreaks[1]
\begin{document}

\hypersetup{pageanchor=false}

\begin{titlepage}

\vspace*{-3.0cm}
\begin{flushright}
\end{flushright}

\begin{center}
{\Large\bf
  Secondary Cosmic Positrons in an Inhomogeneous Diffusion Model
}
\\
\vspace{1cm}
\textbf{
Rolf Kappl,
Annika Reinert
}
\\[5mm]
\textit{\small
Bethe Center for Theoretical Physics \& Physikalisches Institut der 
Universit\"at Bonn, \\
Nu{\ss}allee 12, 53115 Bonn, Germany
}
\end{center}

\vspace{1cm}

\begin{abstract}
One aim of cosmic ray measurements is the search for possible signatures of annihilating or decaying dark matter. The so-called positron excess has attracted a lot of attention in this context. On the other hand it has been proposed that the data might challenge the established diffusion model for cosmic ray propagation. We investigate an inhomogeneous diffusion model by solving the corresponding equations analytically. Depending on the propagation parameters we find that the spectral features of the positron spectrum are affected significantly.
We also discuss the influence of the inhomogeneity on hadronic spectra.
\end{abstract}

\end{titlepage}

\hypersetup{pageanchor=true}

\vspace{2cm}

\section{Introduction}
\label{chap:introduction}

In many particle physics models, dark matter can annihilate or decay into Standard Model particles, which would then propagate in the galaxy and enhance the amount of observed cosmic rays. A major issue for testing this hypothesis is a precise determination of the astrophysical background. Since antiparticles are mainly produced as secondary particles by the spallation of cosmic rays on the interstellar medium (ISM), an estimate of the background contribution is possible. 

Before reaching the Earth, the particles propagate in the Milky Way scattering off the magnetic field, eventually suffering from energy losses, annihilation or escape from the galactic halo. It turns out that the cosmic ray spectra are strongly affected by these processes. An appropriate description for cosmic ray transport is provided by diffusion models \cite{Ginzburg:1990sk,Strong:1998pw,Moskalenko:1997gh,Maurin:2001sj}.
However, cosmic ray propagation is far from being completely understood. It gives rise to the main uncertainties in the derivation of the background and thus demands for more realistic and complex models.

In this light the measured positron flux received significant attention as it exceeds the theoretical predictions for the astrophysical background. This observation was first made by the PAMELA collaboration in 2008 \cite{Adriani:2008zr}, which discovered a rise in the positron fraction $e^+/(e^-+e^+)$. 

Interpretations of this excess consider additional primary sources like dark matter \cite{Bergstrom:2008gr,Cirelli:2008jk}, pulsars \cite{Hooper:2008kg, Linden:2013mqa} or supernova remnants \cite{Fujita:2009wk, Kohri:2015mga}, but the processes responsible for a possible production of positrons are speculative and poorly understood. 
At the experimental side the situation has further improved with the 
recent AMS-02 data \cite{Aguilar:2014mma} which resulted in further 
dark matter studies \cite{Bergstrom:2013jra,Kopp:2013eka,Ibarra:2013zia} 
and background parameterizations including pulsars 
\cite{Gaggero:2013nfa,DiMauro:2014iia,DiMauro:2015jxa}. See also 
\cite{Serpico:2011wg} 
for a review about the positron excess.

An alternative idea was proposed in \cite{Katz:2009yd, Blum:2013zsa, 
Israel:2014mja, Cowsik:2013woa, Blasi:2009hv, Tomassetti:2015mha, 
Dado:2015eda, Ahlen:2014ica, Lipari:2016vqk}, 
stating that a modified diffusion model could explain the observed positron signal from a purely secondary origin. 
An interesting inspiration for this assumption is presented in \cite{Katz:2009yd, Lipari:2016vqk}. Studying the spectrum of cosmic antiprotons which are expected to have (mainly) a secondary origin \cite{Giesen:2015ufa,Evoli:2015vaa,Kappl:2015bqa} an upper bound on the secondary positron spectrum is estimated which agrees with the data. This hypothesis is encouraged by recent AMS-02 data on the antiproton to positron ratio \cite{Aguilar:2016kjl}.

In a model where energy losses are less substantial, for example if the propagation time is reduced, the theoretical prediction for the spectrum would be flatter. This condition can be realized if the diffusion coefficient is taken to increase with galactic height. This approach is motivated by the spatial distribution of the galactic magnetic field which is responsible for the diffusion process. From observations of radio data it is expected to decrease exponentially with distance to the galactic disk (see e.g. \cite{Jansson:2009ip}). 

As a particle propagates, it experiences fewer disturbances for weaker mag\-netic field stren\-gths and prop\-agates freely in the limit of large galactic height $z$. 
In contrast, close to the galactic disk the field lines may capture a charged particle for a while as the direction is frequently changed. 
Hence a decreasing magnetic field corresponds to an increasing diffusion coefficient \cite{Evoli:2008dv,Gebauer:2009hk,Grajek:2010bz}. 

A consequence of the spatial dependence is that particles leaving the vicinity of the disk have a lower probability to come back but will rather drift away. This can be understood recalling the escape time which is anti-proportional to the diffusion coefficient 
\cite{Ginzburg:1990sk}. With increasing galactic height $z$ the escape time gets smaller, signifying the decreasing probability of a particle to return to the disk. 
In opposition to a homogeneous and isotropic diffusion coefficient this implies that particles detected in the disk are less likely to originate from a source which is far away. In other words the radius from where particles reach the Earth is reduced and energy losses become less important.
This does not spoil the constraints from secondary to primary ratios of cosmic nuclei as their spectrum is nearly insensitive to the propagation time since energy losses have a much smaller impact on hadronic spectra \cite{TalkBlum}.

Some special cases of inhomogeneous diffusion can be found in the literature. In the \textsc{Dragon}-package \cite{Evoli:2008dv} a vertically, exponentially increasing diffusion coefficient is implemented. An updated version also takes the spiral structure of the source distribution into account \cite{Evoli:2016xgn}. 
Another discussion is presented in \cite{Gebauer:2009hk} where a diffusion coefficient increasing linearly in vertical direction is implemented into the \textsc{Galprop}-code.  
A similar setting is investigated in \cite{Grajek:2010bz} studying the influence of the same modification on the antiproton spectrum. In these works it is pointed out that a diffusion model with vertically increasing diffusion coefficient is compatible with B/C and $^{10}\text{Be}/^{9}\text{Be}$ measurements.

All these approaches focus on a numerical treatment, whereas we present analytic solutions for different kinds of vertical inhomogeneous diffusion models for leptons and hadrons\footnote{Some analytical results for special cases of inhomogeneous diffusion are discussed in the literature (see e.g. \cite{1975ICRC....2..706B,1980ApJ...239.1089L}).}. Concretely we consider a diffusion coefficient with an arbitrary power-law and exponential dependence in the vertical coordinate $z$.

The paper is organized as follows. In chapter \ref{chap:isotropicDiffusion} we review the isotropic two-zone diffusion model. In chapter \ref{chap:spaceDepDiffusion} we present an analytic solution to the inhomogeneous diffusion equation for positrons and hadrons. We show that the high energy part of the positron spectrum and the B/C ratio can be reproduced in this framework. After a short discussion on the propagation parameters we conclude in section \ref{chap:conclusion}. All details of the calculation are sketched in appendix \ref{app:solve}.

\section{The Isotropic Two-Zone Diffusion Model}
\label{chap:isotropicDiffusion}

The main idea of the two-zone diffusion model \cite{Ginzburg:1990sk,Strong:1998pw,Moskalenko:1997gh,Maurin:2001sj} is the separation of a source region and a homogeneous, isotropic diffusion halo. In the following we call this model isotropic model. 
The galaxy is described by a thin disk of radius $R=\SI{20}{kpc}$ and half-thickness $h=\SI{0.1}{kpc}$ containing the sources like stars, supernovae and cosmic rays scattering off the ISM. The disk residing at $z=0$ is surrounded by a cylindrical halo with half-height $L$. The solar system is located in the disk at a radial distance $R_0 = \SI{8.5}{kpc}$ to the galactic center. 

In the standard model of cosmic ray propagation the stationary diffusion equation for the particle density $\mathcal{N}(E,\mathbf{x})$ reads \cite{Putze:2010zn}
\begin{equation}
\label{eq:DiffusionGeneral}
\nabla\left(-K\nabla\mathcal{N}+\textbf{V}_c\mathcal{N}\right)+\frac{\partial}{\partial E}\left(b^{\text{tot}}\mathcal{N}-\beta^2K_{EE}\frac{\partial}{\partial E}\mathcal{N}\right)+2h\delta(z)\Gamma^{\text{ann}}\mathcal{N}=Q.
\end{equation}
The energy dependent diffusion coefficient $K$ and the diffusion coefficient in energy space $K_{EE}$ responsible for reacceleration are \cite{Strong:1998pw, Putze:2010zn}
\begin{equation}
\label{eq:reaccelerationCoeff}
K=K_0\beta\mathcal{R}^\delta, \qquad K_{EE}=\frac{4}{3}\frac{V_a^2}{K}\frac{p^2}{\delta(4-\delta^2)(4-\delta)}
\end{equation}
where $K_0$ is the diffusion constant, $\mathcal{R}$ rigidity, $\delta$ the index of the rigidity dependent slope, $p$ momentum and $V_a$ the Alfv\`{e}n velocity.
The convective wind velocity is perpendicular to the galactic plane $\mathbf{V}_c=V_c\hat{\textbf{e}}_z$ and annihilation $\Gamma^{\text{ann}}$ takes place only in the interstellar disk. \(\beta\) denotes the velocity in units of \(c\). The energy loss term $b^{\text{tot}}$ contains losses of all kinds and a term induced by reacceleration
\begin{equation}
b^{\text{tot}}=b^{\text{loss}}+\frac{1+\beta^2}{E}K_{EE}\, .
\end{equation}
For the sake of simplicity we do not consider convection and reacceleration in the inhomogeneous model discussed in this work.
The source term \(Q(E,\mathbf{x})\) 
contains primary and secondary cosmic ray sources.
The particle number density $\mathcal{N}$ is related to the flux $\Phi$ at earth via
\begin{equation}
\label{eq:flux}
\Phi(E)=\frac{\beta}{4\pi}\mathcal{N}(E,R=R_0, z=0)\, .
\end{equation}
At the spatial boundaries Dirichlet conditions are imposed, signifying that particles reaching the edge of the galactic halo escape into outer space. The propagation model depends on five free parameters $L, K_0, \delta, V_c$ and $V_a$ which have to be determined experimentally. We restrict ourselves to one spatial dimension \(z\), perpendicular to the galactic plane. As the propagation length does not exceed a few kpc, we can safely ignore the radial boundaries (see e.g. \cite{Genolini:2015cta}).

For positrons the diffusion equation can be solved analytically if only diffusion and energy losses are considered\footnote{Some analytical results including convection are also available in the literature \cite{1980ApJ...239.1089L}.}.
The two main approaches are based on the use of the image method for Green's functions or an eigenfunction expansion (see e.g. \cite{1974Ap&SS..29..305B,Baltz:1998xv,Delahaye:2008ua,Delahaye:2010ji}).
For a general source distribution we have
\begin{equation}
\mathcal{N}(E,z)=\intop_E^\infty \dd E_s\intop_{-\infty}^\infty \dd z_s \mathcal{G}(E,E_s,z,z_s)Q(E_s,z_s).
\end{equation}
In the case of positrons the diffusion equation can be simplified to
\begin{equation}
\label{eq:posAni}
-\frac{\partial}{\partial z}\left[K\frac{\partial}{\partial z}\mathcal{N}\right]+\frac{\partial}{\partial E}\left(b^{\text{loss}}\mathcal{N}\right)=Q,\qquad b^{\text{loss}}=-b_0E^2
\end{equation}
with \(b_0=10^{-16} \unit{s^{-1}}\) accounting only for diffusion and energy losses due to synchrotron radiation and inverse Compton scattering \cite{Delahaye:2008ua}.
We further set \(\mathcal{R}=p=E\) as the mass of positrons with respect to their kinetic energy can be neglected. It is convenient to transform the energy \(E\) to a new variable
\begin{equation}
\lambda=-\intop_E^\infty\dd E^\prime\frac{K(E^\prime)}{\big\vert b(E^\prime)\big\vert}=\frac{K_0}{b_0}
\frac{E^{\delta-1}}{1-\delta}\in [\infty, 0]\, .
\end{equation}
If the diffusion coefficient is assumed to be isotropic we can use the Green's function from \cite{Lavalle:2006vb,Delahaye:2008ua}
\begin{equation}
\mathcal{G}(\lambda,\lambda_s,z,z_s)=\frac{2}{L}\sum_{n=1}^{\infty} e^{-k_n^2(\lambda-\lambda_s)}\cos(k_n z)\cos(k_n z_s)\, ,\qquad k_n=\left(n-\frac{1}{2}\right)\frac{\pi}{L}
\end{equation}
where we took only the even part in \(z\) for symmetry reasons. Assuming that the sources are homogeneously distributed in the galactic disk with half-height \(h\) like
\begin{equation}
\label{eq:sourceDisk}
Q(E,z)=\begin{cases}
q_0E^{-\gamma_0}\, ,&\qquad -h < z < h\\
0\, ,&\qquad \text{elsewhere}
\end{cases}
\end{equation}
we recover the result from \cite{1974Ap&SS..29..305B}
\begin{equation}
\label{eq:BD}
\begin{split}
\mathcal{N}(E,z)&=\frac{2q_0}{b_0}\frac{E^{-(\gamma_0+1)}}{\gamma_0-1}\sum_{n=0}^\infty\frac{\cos\left(\pi\left(n+\frac{1}{2}\right)\frac{z}{L}\right)\sin\left(\pi\left(n+\frac{1}{2}\right)\frac{h}{L}\right)}{\pi(n+\frac{1}{2})}\\
&\phantom{=}\cdot {}_1F_1\left(1,\frac{\gamma_0-\delta}{1-\delta};-\frac{\pi^2\left(n+\frac{1}{2}\right)^2}{L^2}\frac{K_0}{b_0}\frac{E^{\delta-1}}{1-\delta}\right)\, .
\end{split}
\end{equation}
The energy integration results in the occurrence of the confluent hypergeometric function. 
\begin{figure}
\centering
\includegraphics[width=0.7\textwidth]{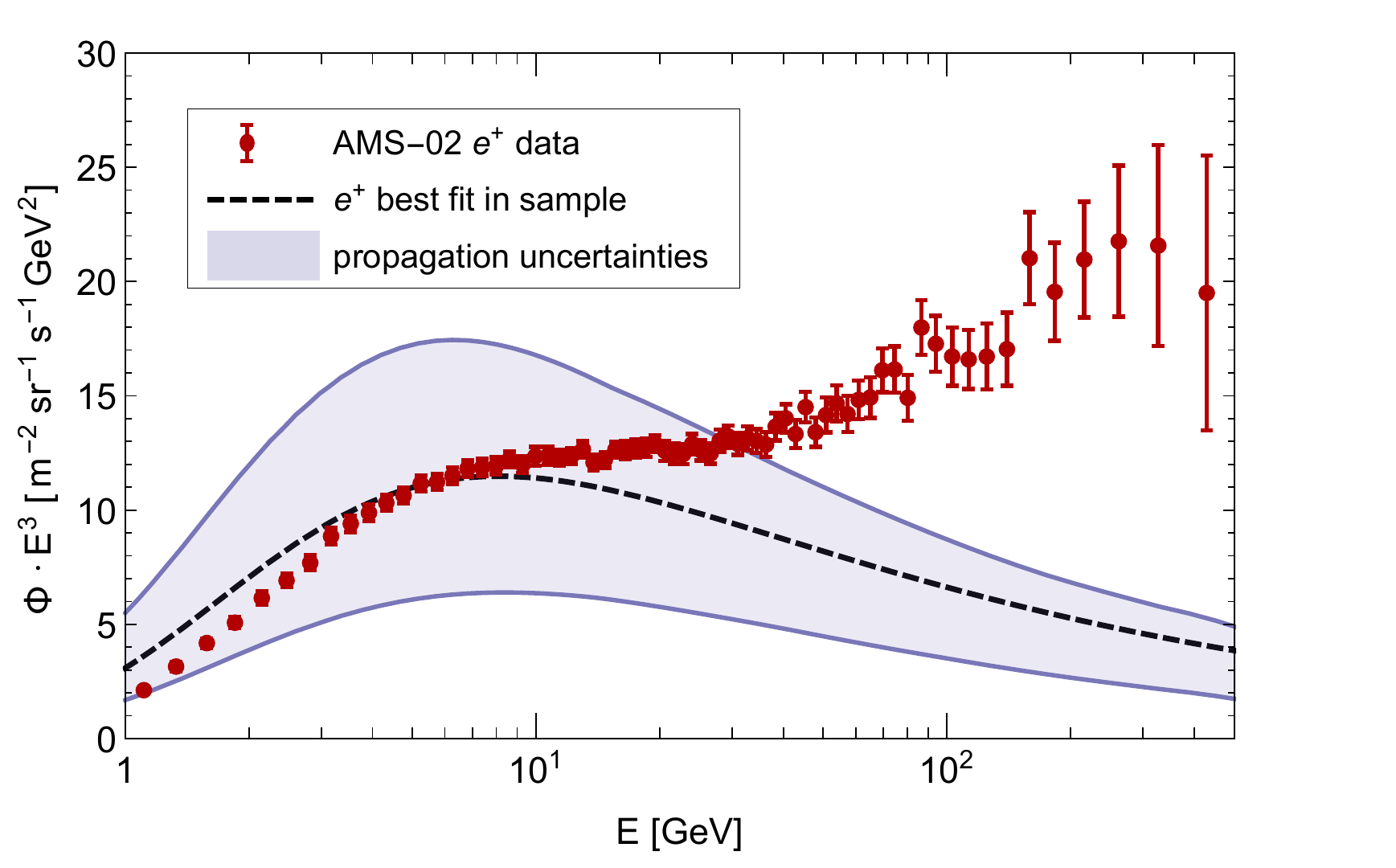}
\caption{The secondary positron spectrum predicted in the two-zone diffusion model compared to the data from AMS-02 \cite{Aguilar:2014mma}. The inner band encompasses propagation uncertainties (see text) and the dashed line indicates the configuration from our sample which best fits the data up to $\sim\SI{20}{\GeV}$. Respecting solar modulation the theoretical curves were modified with a Fisk potential of $\phi=\SI{0.57}{GV}$.}
\label{fig:positronsDM}
\end{figure}
Relevant for observations is the flux at the position of the earth, which is given by
\begin{equation}
\label{eq:BDref}
\Phi(E)=\frac{\beta}{4\pi}\frac{2h}{L}\frac{q_0}{b_0}\frac{E^{-(\gamma_0+1)}}{\gamma_0-1}\sum_{n=1}^{\infty}\frac{\sin \left(k_n h\right)}{k_n h}
{}_1F_1\left(1,\frac{\gamma_0-\delta}{1-\delta};
-k_n^2\frac{K_0}{b_0}\frac{E^{\delta-1}}{1-\delta}\right).
\end{equation}

Let us shortly comment on the source term for positrons.
A detailed discussion on the derivation of the secondary positron source term can be found in \cite{Delahaye:2008ua}. We recomputed the results numerically with the cross sections from \cite{Kamae:2006bf} and use a simplified expression for the analytical treatment\footnote{An improvement for the cross sections using recent NA49 data\cite{Alt:2005zq} would be desirable like it was achieved for the antiproton production cross sections\cite{Kappl:2014hha,diMauro:2014zea}.}. 
This expression is given by
\begin{equation}
\label{eq:sourcetermPos}
Q(E,z)=2hq_0\delta(z)E^{-\gamma_0}
\end{equation} 
with $q_0\approx 5\cdot 10^{-27}\unit{cm^{-3}s^{-1}GeV^{-1}}$ and $\gamma_0\approx 2.65$ \cite{Cowsik:2013woa}.
The flux at earth for this source term can be found to be
\begin{equation}
\label{eq:BDphi}
\Phi(E)=\frac{\beta}{4\pi}\frac{2h}{L}\frac{q_0}{b_0}\frac{E^{-(\gamma_0+1)}}{\gamma_0-1}\sum_{n=1}^{\infty}
{}_1F_1\left(1,\frac{\gamma_0-\delta}{1-\delta};
-k_n^2\frac{K_0}{b_0}\frac{E^{\delta-1}}{1-\delta}\right)\, ,
\end{equation}
which can be obtained from \eqref{eq:BDref} by taking the limit \(h\ll L\).

For free boundaries (this is a viable assumption for energies exceeding a few $\GeV$) a simple energy scaling can be found for $\Phi$.
It can be shown that 
(see e.g. \cite{1980ApJ...239.1089L,Delahaye:2010ji})
\begin{equation}
\label{eq:exponentconst}
\Phi(E)\propto E^{-\gamma}=E^{-\gamma_0-\frac{1}{2}(1+\delta)}.
\end{equation}

We compare the full result for \(\Phi\) to the data from AMS-02 \cite{Aguilar:2014mma} in figure \ref{fig:positronsDM}. The propagation uncertainties are estimated by enveloping 500 sets of propagation parameters which are consistent with the most recent boron to carbon data \cite{Kappl:2015bqa} (we neglect convection and reacceleration in figure \ref{fig:positronsDM}). The configuration that best fits the data up to $\sim \SI{20}\GeV$ is indicated by the dashed line. For solar modulation we use the force-field approximation \cite{Gleeson:1968zza} because a more detailed analysis for leptons is challenging (see e.g. \cite{Potgieter:2015jxa,Kappl:2015hxv}).

Although the data are tolerably described in the low energy regime there is a gap between the prediction from the two-zone diffusion model and the data above $\sim \SI{30}{GeV}$ which is known as the positron excess. In the next section we investigate a phenomenologically well motivated modification of the diffusion model in order to find a description for the positron data in the high energy regime.

\section{Inhomogeneous Diffusion}
\label{chap:spaceDepDiffusion}

In this section we present an analytic solution to the inhomogeneous diffusion model including energy losses and its impact on the cosmic ray spectra. Following an analytic approach, convection and reacceleration are neglected in this work. The latter can be included in a second step in a semi-analytical approach, similar as in \cite{Donato:2001ms}. The existence of convective winds is controversially discussed \cite{Strong:2007nh} and is neglected in order to reduce the amount of parameters. This assumption is consistent with recent B/C analyses (see table \ref{tab:bestBC}).

\begin{table}
\centering
\begin{tabular}{cccccc}
\(\delta\)&\(K_0\ (\text{kpc}^2\ \text{Myr}^{-1})\)
&\(L\ (\text{kpc})\)&\(V_c\ (\text{km}\ \text{s}^{-1})\)&\(V_a\ 
(\text{km}\ \text{s}^{-1})\)&\(\chi^2\ /\text{d.o.f.}\)\\
\hline
0.408&0.0967&13.7&0.2&31.9&1.19\\
\end{tabular}
\caption{Propagation parameters which best fit the preliminary data of the 
B/C ratio from the AMS-02 collaboration as derived in \cite{Kappl:2015bqa}.}
\label{tab:bestBC}
\end{table}

\subsection{Propagation of Positrons}

As motivated in the introduction we consider a spatial dependent diffusion coefficient 
\begin{equation}
\label{eq:DiffCoefficient}
K\mapsto K(E,z)=K_0E^\delta f(z)
\end{equation}
depending on energy $E$ with $0\leq\delta<1$ and a general function of the spatial coordinate $z$.

The diffusion equation \eqref{eq:posAni} is solved for two different functional dependencies of the diffusion coefficient \eqref{eq:DiffCoefficient} on the galactic height $z$. Details on the derivation are provided in appendix \ref{app:solve}. 
For $f(z)=|z|^\mu$ we find the solution for the flux at earth
\begin{equation}
\label{eq:NumberDensityPwl}
\begin{split}
\Phi(E)&=\frac{\beta}{4\pi}\frac{2}{\left|2-\mu\right|}\frac{2h}{L}\frac{q_0}{b_0}\frac{E^{-(\gamma_0+1)}}{\gamma_0-1}
\sum_{n=1}^{\infty} 
\frac{{}_0F_1\left(\frac{1}{2-\mu};-\frac{1}{4}\zeta_n^2\left(\frac{\xi}{L}\right)^{2-\mu}\right)^2}
{\zeta_n^2\ {}_0F_1\left(1+\frac{1}{2-\mu};-\frac{1}{4}\zeta_n^2\right)^2}\\
&\phantom{=}\cdot 
{}_1F_1\left(1,\frac{\gamma_0-\delta}{1-\delta};
-\Omega_n^2\frac{K_0}{b_0}\frac{E^{\delta-1}}{1-\delta}\right)
\, ,
\end{split}
\end{equation}
with $\Omega_n=\frac{2-\mu}{2}\zeta_n L^{-\frac{2-\mu}{2}}$. $_0F_1$ denotes the confluent hypergeometric limit function and $_1F_1$ is a confluent hypergeometric function like in the isotropic case. $\zeta_n$ denote the zeros of the Bessel function of the first kind and order \(-\frac{1-\mu}{2-\mu}\). 
We have further introduced a small cutoff \(\xi\) to ensure that the diffusion coefficient \(K(E,z)\) is non-vanishing in the galactic disk.
This is discussed in more detail in appendix \ref{app:powerlaw}. \(\xi\) can be understood as a parameter determining the diffusion strength in the galactic plane. We should also point out, that in this model diffusion is smaller than in the isotropic model in the region \(z < 1\) and stronger for higher latitudes \(z>1\).

The functional form of the inhomogeneous result is very similar to the isotropic one. In the hypergeometric function, \(k_n\) is replaced by \(\Omega_n\) and the sine and cosine are replaced by the occurrence of the hypergeometric limit function. The isotropic result can be obtained exactly by taking the spatial dependence \(\mu\) to zero as shown in appendix \ref{app:isotropic}.

The solution for an exponentially increasing diffusion coefficient $f(z)=\text{e}^{az}$ with an arbitrary factor $a$ is found to be
\begin{equation}
\label{eq:NumberDensityExp}
\Phi(E)=\frac{\beta}{4\pi}2h\frac{q_0}{b_0}|a|e^{aL}
\frac{E^{-(\gamma_0+1)}}{\gamma_0-1}
\sum_{n=1}^\infty\left[
\frac{\mathcal{J}_1\left(\zeta_ne^{\frac{a}{2}L}\right)}
{\mathcal{J}_2(\zeta_n)}
\right]^2
{}_1F_1\left(1,\frac{\gamma_0-\delta}{1-\delta};-\Lambda_n^2\frac{K_0}{b_0}\frac{E^{\delta-1}}{1-\delta}\right)\, ,
\end{equation}
with $\Lambda_n=\frac{a}{2}\zeta_n e^{\frac{a}{2}L}$.

In the following we want to explore the spectral features of the new models for $\mu, a>0$. To do so we make use of the behavior of the confluent hypergeometric function in the limits \cite{abramowitz} 
\begin{equation}
{}_1F_1(\alpha,\beta;-x)\xrightarrow{x\to\infty} \frac{\Gamma(\beta)}{\Gamma(\beta-\alpha)}x^{-\alpha}
\qquad \text{and}\qquad {}_1F_1(\alpha,\beta;-x)\xrightarrow{x\to 0} 1\ .
\end{equation}
Since $\delta<1$ is evident from secondary to primary ratios \cite{Maurin:2001sj} the exponent \(\delta-1\) of the energy $E$ in the hypergeometric function is negative. Consequently in the low and high energy regime the flux is proportional to\footnote{In general each term of the sum over $n$ has to be treated separately for the analysis of the spectral behavior. Here the coefficient in the hypergeometric function increases with higher order so the low energy regime is defined by the term of lowest order, i.e.\! $n=1$.} 
\begin{equation}
\Phi(E)\propto 
\begin{cases}
E^{-(\gamma_0+\delta)}&,\text{ low energies}\\
E^{-(\gamma_0+1)}&,\text{ high energies}\, .
\end{cases}
\end{equation} 
These results agree with the description of different regimes given in \cite{1974Ap&SS..29..305B} for the isotropic case. 

Let us comment on the high energy behavior of positrons and the implications for the positron excess.
The high energy part of the positron spectrum measured at the Earth has a spectral index of about $\gamma=2.75$.
There is only a small difference to the spectral index of the source term $\gamma_0$ of roughly $\Delta\gamma_\text{obs}=\gamma-\gamma_0\lesssim 0.1$. In the inhomogeneous diffusion model this can be realized for a small value of the spectral slope $\delta$ if we are still in the low energy regime since we find $\Delta\gamma_\text{inhomogeneous}=\delta$. Hence for small values for $\delta$ the positron spectrum can be fitted 
if the turnover between the low energy and the high energy regime is shifted by adjusting the parameters in the argument of the hypergeometric function accordingly 
up to positron energies of $\sim\SI{400}{GeV}$. 

Contrarily, in the isotropic two-zone diffusion model this is impossible for reasonable parameters since $\Delta\gamma_\text{isotropic}=\frac{1+\delta}{2}\geq 0.5$, as derived in \eqref{eq:exponentconst}.
For very small \(L\) the same behavior as in the inhomogeneous model could in principle be found which can be seen by inspection of \(k_n\). 
Comparing the solution for isotropic diffusion \eqref{eq:BDref} with the inhomgeneous cases \eqref{eq:NumberDensityPwl} and \eqref{eq:NumberDensityExp} we find that they have a very similar structure. Concerning the energy dependence, the only difference lies within one characteristic factor, namely \(k_n\), \(\Omega_n\) or \(\Lambda_n\) in the hypergeometric function and its weights in the sum. We have
\begin{equation}
k_n=\left(n-\frac{1}{2}\right)\frac{\pi}{L},\qquad \Omega_n=\frac{2-\mu}{2}\zeta_n L^{-\frac{2-\mu}{2}},\qquad \Lambda_n=\frac{a}{2}\zeta_n e^{\frac{a}{2}L}\, .
\end{equation}
The energy dependence is dominated by an overall factor of $E^{-(\gamma_0+1)}$ which gets corrections from the hypergeometric function. If \(L\) is fixed, we have no further freedom to adjust \(k_n\) and the isotropic behavior is fixed. In the inhomogeneous cases, \(\mu\) and \(a\) allow for a tuning and we can use it to shift the turnover to energies that are high enough to explain the high energy behavior of cosmic positrons.

\begin{figure}
	\centering
	\includegraphics[width=0.49\textwidth]{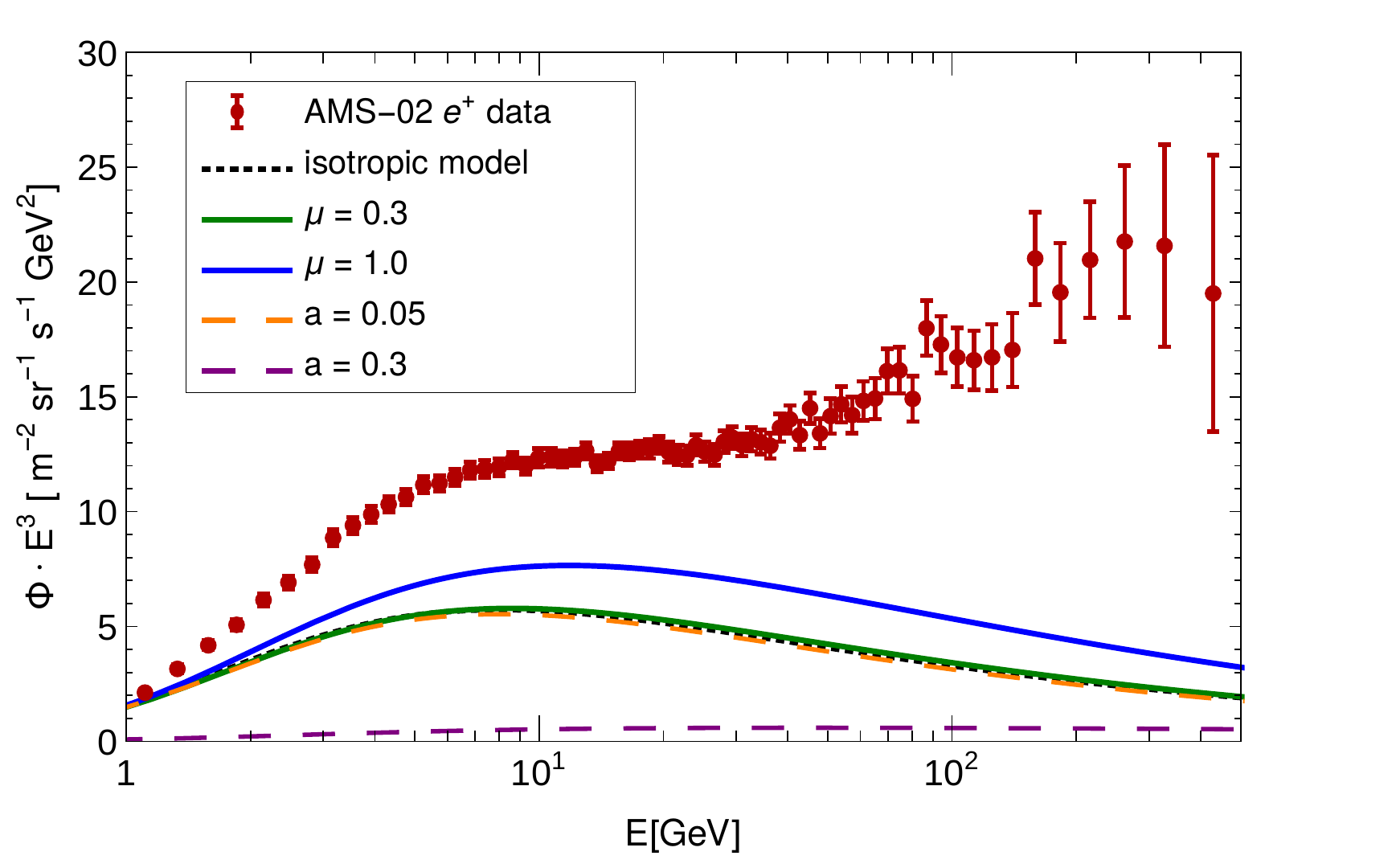}
	\includegraphics[width=0.49\textwidth]{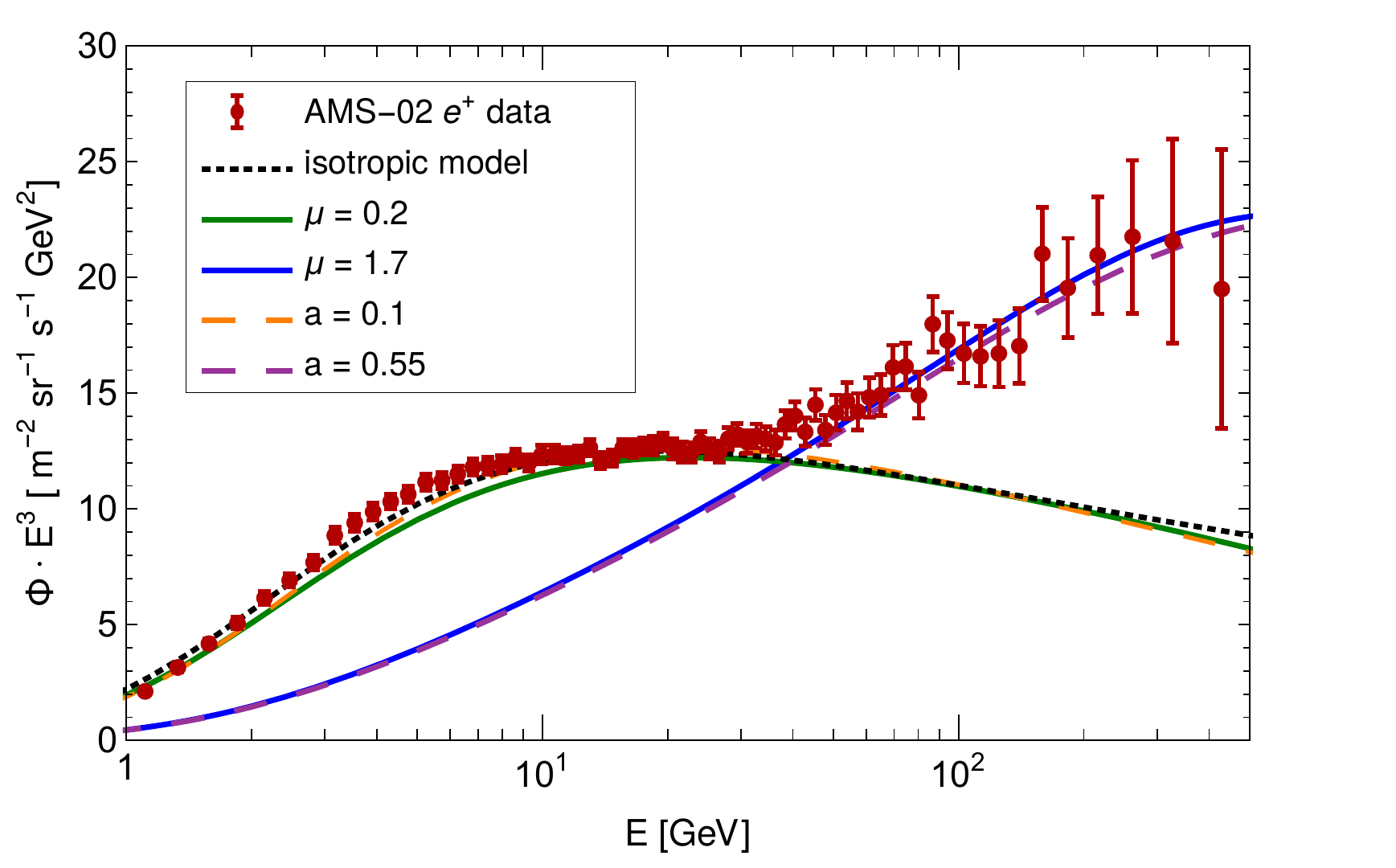}
	\caption{Left: The positron spectrum measured by AMS-02 with the secondary background derived in a diffusion model with spatial dependent diffusion coefficient, increasing in $z$ with a power-law (solid lines) and an exponential (dashed lines) dependence. The functions were modulated with a force field $\phi=\SI{0.57}{GV}$ accounting for solar modulation. The diffusion parameters are given in the legend and table \ref{tab:bestBC}. Right: The same plot but with parameters given in table \ref{tab:propParameters}.}
	\label{fig:positronsnew}
\end{figure}

\begin{table}
\centering
\begin{tabular}{lcccccc}
\(\text{color}\)&\(K_0\ (\text{kpc}^2\ \text{Myr}^{-1})\)&\(\mu\)&\(\xi\ (\text{kpc}) \)&\(a\ (\text{kpc}^{-1})\)\\
\midrule
\text{black}&0.06&-&-&-\\
\text{green}&0.08&0.2&0.01&-\\
\text{blue}&3.5&1.7&0.01&-\\
\text{orange}&0.05&-&-&0.1\\
\text{purple}&0.005&-&-&0.55\\
\end{tabular}
\caption{The propagation parameters used in figure \ref{fig:positronsnew} and \ref{fig:1F1}. In all cases we set $L=10 \text{ kpc}$ and $\delta=0$.}
\label{tab:propParameters}
\end{table}

Under these considerations we present different configurations of the new model with small values for $\delta$ in figure \ref{fig:positronsnew}, properly fitting the positron data in the high energy regime. We also present the prediction from the isotropic model with the same value for the spectral slope. For the sake of comparison with the other models the diffusion constant was chosen freely, such that it best fits the positron data to energies as high as possible.
Note that a small spectral slope $\delta$ leads to problems in reproducing the B/C ratio. However, proposes for a vanishing $\delta$ for positrons can be found in the literature \cite{Cowsik:2013woa}. Vanishing \(\delta\) is also in agreement with the recent antiproton data from AMS-02 \cite{Aguilar:2016kjl}. 
We can also see in the left plot of figure \ref{fig:positronsnew} the outlined feature that in the power-law model, parametrized by \(\mu\), the amplitude increases due to the reduced value of the diffusion coefficient in the galactic disk.

\begin{figure}
	\centering
	\includegraphics[width=0.49\textwidth]{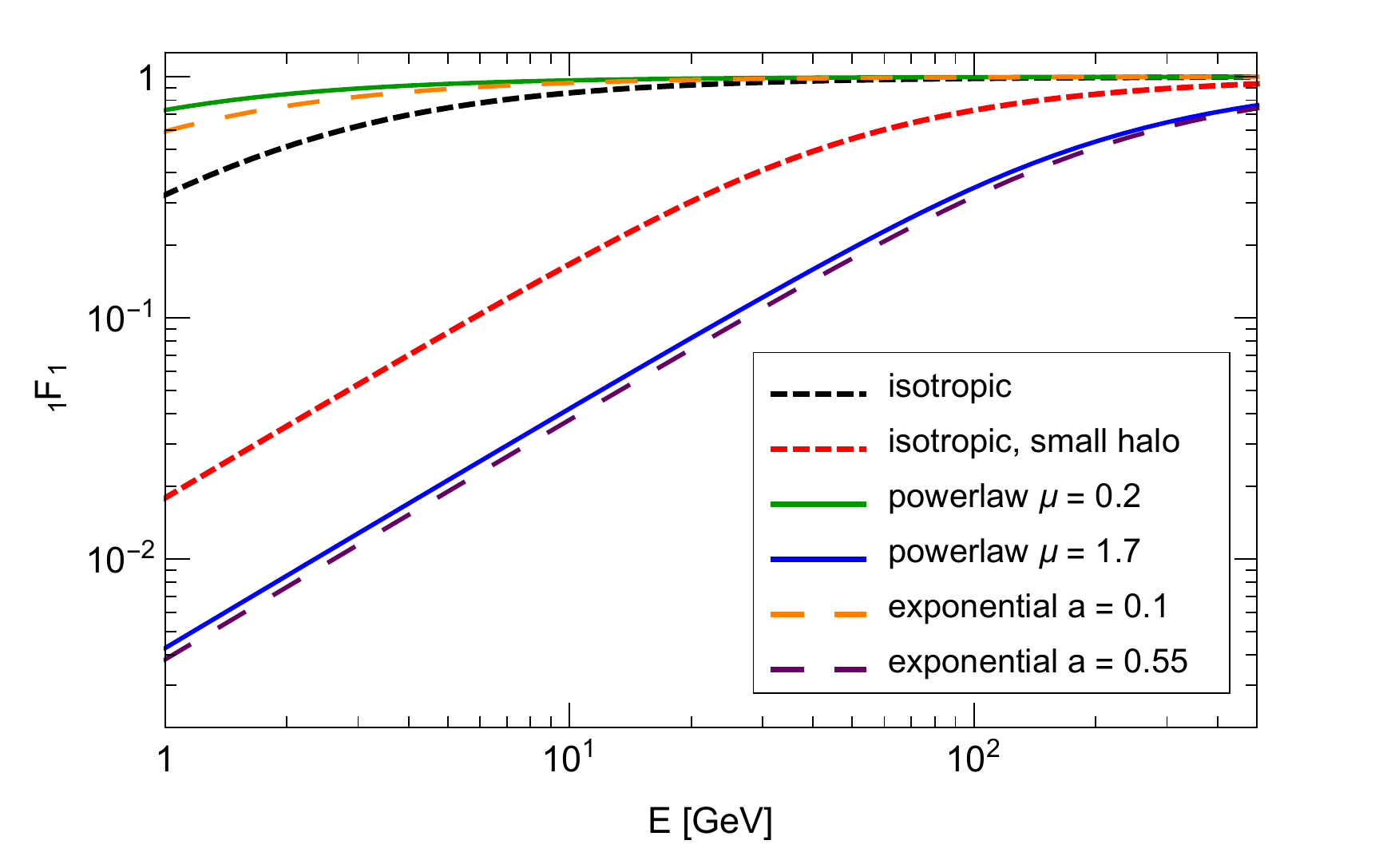}
	\caption{The energy dependence of the hypergeometric functions appearing in the equations for the lepton flux \eqref{eq:BDref}, \eqref{eq:NumberDensityPwl} and \eqref{eq:NumberDensityExp} with the propagation parameters used in figure \ref{fig:positronsnew}, given in table \ref{tab:propParameters}. Additionally shown is the isotropic case for $L=1 \text{ kpc}$ (red dotted line).}
	\label{fig:1F1}
\end{figure}

The influence of the spatial dependence is explained in figure \ref{fig:1F1}, where the hypergeometric functions for the propagation parameters used in figure \ref{fig:positronsnew} are shown.
One can see that in the isotropic case the hypergeometric function is nearly constant. The same holds for small inhomogeneities, hence for all propagation parameter sets which do not suffice to fit the high energy part of the positron spectrum. 
For larger inhomogeneities the curve changes, resulting in a harder spectrum for energies up to some hundreds of GeV. This feature gives rise to the differences observed in the spectra in figure \ref{fig:positronsnew}. We also demonstrate that in an isotropic diffusion model with small halo a similar effect is achieved, as also discussed in \cite{Delahaye:2008ua}.

We also want to comment briefly on the choice of the propagation parameters in table \ref{tab:propParameters}. One can see from equation \eqref{eq:NumberDensityPwl} and \eqref{eq:NumberDensityExp} 
that the position of the turnover in the spectrum depends strongly on $K_0$ and also on $\mu$ or \(a\). Thus for each model of inhomogeneity a specific value for $K_0$ can be chosen such that the turnover is located at about $\SI{400}{GeV}$. 

As clearly visible in figure \ref{fig:positronsnew} our inhomogeneous model is unable to describe the positron spectrum over the whole energy range. Fitting the high energy part of the positron spectrum a gap appears in the low energy part. One possibility to cure this problem is given by including reacceleration, which is known to give additional contributions to the low energy part of cosmic ray spectra. This is not possible in an analytical way, at least none that we are aware of. We tried to include reacceleration in a semi-analytical framework, but it was impossible to cure the outlined problem of the inhomgeneous model. Another origin of the spectral break in the positron spectrum could rely on the spectral break of proton and helium spectra, observed around $\SI{400}{GeV/n}$. As secondary positrons adopt only a fraction of their progenitor's energy, the break could translate into the positron spectrum at energies in the tens of GeV range.

\subsection{Propagation of Hadrons}
\label{chap:anisohad}

Having found a propagation model potentially reproducing the high energy part of the positron spectrum from a secondary origin, we want to check its impact on cosmic nuclei.
For the propagation of hadrons energy losses are less substantial but elastic and inelastic interactions have to be taken into account. Similarly as presented in \cite{Donato:2001ms,Maurin:2001sj}, the spatial parts as well as the energy dependent part of the particle density can be separated and solved. 
For spatial dependent diffusion we start with
\begin{equation}
\label{eq:hadronsDiffEqHighEnergy}
-K_0\beta\mathcal{R}^\delta\frac{\del}{\del z}\left[f(z)\frac{\del}{\del z}\mathcal{N}\right]+2h\delta(z-\xi)\Gamma^{\text{ann}}\mathcal{N}=2hq_0\delta(z-\xi)\mathcal{R}^{-\gamma_0}\, .
\end{equation}
The cutoff \(\xi\) ensures that the diffusion coefficient in the galactic plane is non-vanishing. Introducing 
\begin{equation}
\tilde{q}(E)=\frac{2h}{\beta K_0}\left(q_0\mathcal{R}^{-(\gamma_0+\delta)}-\Gamma^\text{ann}\mathcal{R}^{-\delta}\mathcal{N}(E,\xi)\right)\, ,
\end{equation}
equation \eqref{eq:hadronsDiffEqHighEnergy} can be simplified to
\begin{equation}
-\frac{\del}{\del z}\left[f(z)\frac{\del}{\del z}\mathcal{N}\right]=\tilde{q}(E)\delta(z-\xi)\, .
\end{equation}
As low energy effects like energy losses and reacceleration are neglected, we find an ordinary differential equation.
Applying the leptonic result (one can use \eqref{eq:Greensfunction} setting \(s=0\)) to the power-law case of spatial dependence \(f(z)=|z|^{\mu}\) results in 
\begin{equation}
\begin{split}
\mathcal{N}(E,z)&=
\frac{2}{\left|2-\mu\right|}\frac{1}{L}
\sum_{n=1}^{\infty} 
\frac{{}_0F_1\left(\frac{1}{2-\mu};-\frac{1}{4}\zeta_n^2\left(\frac{\xi}{L}\right)^{2-\mu}\right){}_0F_1\left(\frac{1}{2-\mu};-\frac{1}{4}\zeta_n^2\left(\frac{z}{L}\right)^{2-\mu}\right)}
{\Omega_n^2\zeta_n^2\ {}_0F_1\left(1+\frac{1}{2-\mu};-\frac{1}{4}\zeta_n^2\right)^2}\\
&\phantom{=}\cdot
\frac{2h}{\beta K_0}\left(q_0\mathcal{R}^{-(\gamma_0+\delta)}-\Gamma^\text{ann}\mathcal{R}^{-\delta}\mathcal{N}(E,\xi)\right).
\end{split}
\end{equation} 
We can find \(\Phi\) by setting \(z=\xi\) and solving for \(\mathcal{N}(E,\xi)\).
For an exponential dependence of the diffusion coefficient no cutoff is needed and the result is
\begin{equation}
\mathcal{N}(E,z)=
\frac{4}{|a|} e^{a\left(L-\frac{z}{2}\right)}\sum_{n=1}^{\infty}\frac{\mathcal{J}_1\left(\zeta_n e^{\frac{a}{2}L}\right)\mathcal{J}_1\left(\zeta_ne^{\frac{a}{2}(L-z)}\right)}{\zeta_n^2\left[\mathcal{J}_2(\zeta_n)\right]^2}
\frac{2h}{\beta K_0}\left(q_0\mathcal{R}^{-(\gamma_0+\delta)}-\Gamma^\text{ann}\mathcal{R}^{-\delta}\mathcal{N}(E,0)\right).
\end{equation}
For the comparison with the isotropic case let us neglect interactions with the ISM and set \(\Gamma^\text{ann}=0\) and \(z=\xi\) or \(z=0\), respectively. We find
\begin{equation}
\Phi(E)=\frac{1}{4\pi}\frac{2h}{L}\frac{2}{|2-\mu|}\frac{q_0}{K_0}\mathcal{R}^{-(\gamma_0+\delta)}
\sum_{n=1}^{\infty}
\frac{{}_0F_1\left(\frac{1}{2-\mu};-\frac{1}{4}\zeta_n^2\left(\frac{\xi}{L}\right)^{2-\mu}\right)^2}
{\Omega_n^2\zeta_n^2\ {}_0F_1\left(1+\frac{1}{2-\mu};-\frac{1}{4}\zeta_n^2\right)^2}
\end{equation}
and
\begin{equation}
\Phi(E)=\frac{1}{4\pi}
\frac{8h}{|a|}\frac{q_0}{K_0}\mathcal{R}^{-(\gamma_0+\delta)}\sum_{n=1}^{\infty}\left[
\frac{\mathcal{J}_1\left(\zeta_ne^{\frac{aL}{2}}\right)}{\zeta_n\mathcal{J}_2(\zeta_n)}\right]^2
\end{equation}
for the two different cases of inhomogeneity.

\begin{figure}
	\centering
	\includegraphics[width=0.48\textwidth]{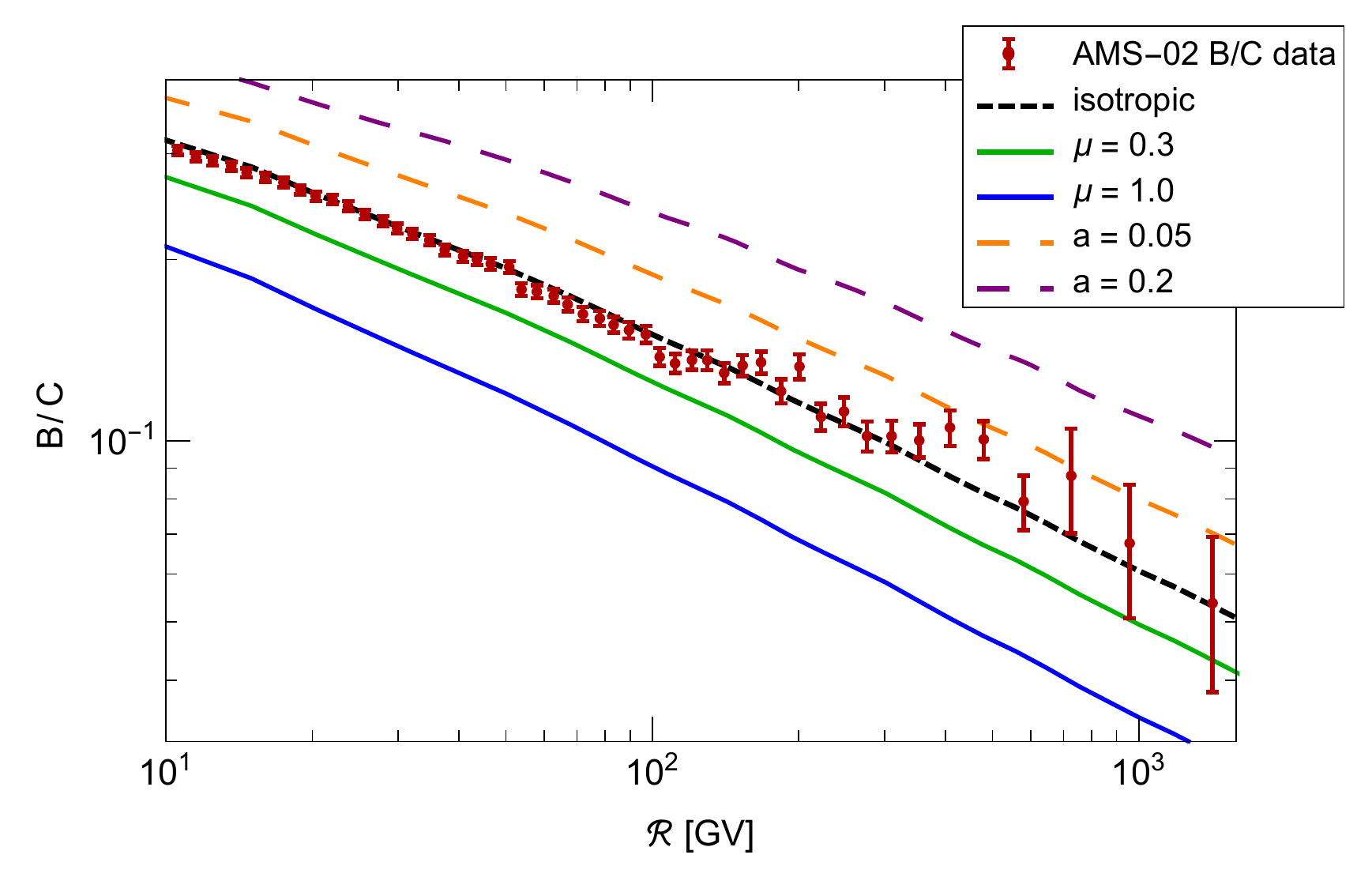}
	\includegraphics[width=0.51\textwidth]{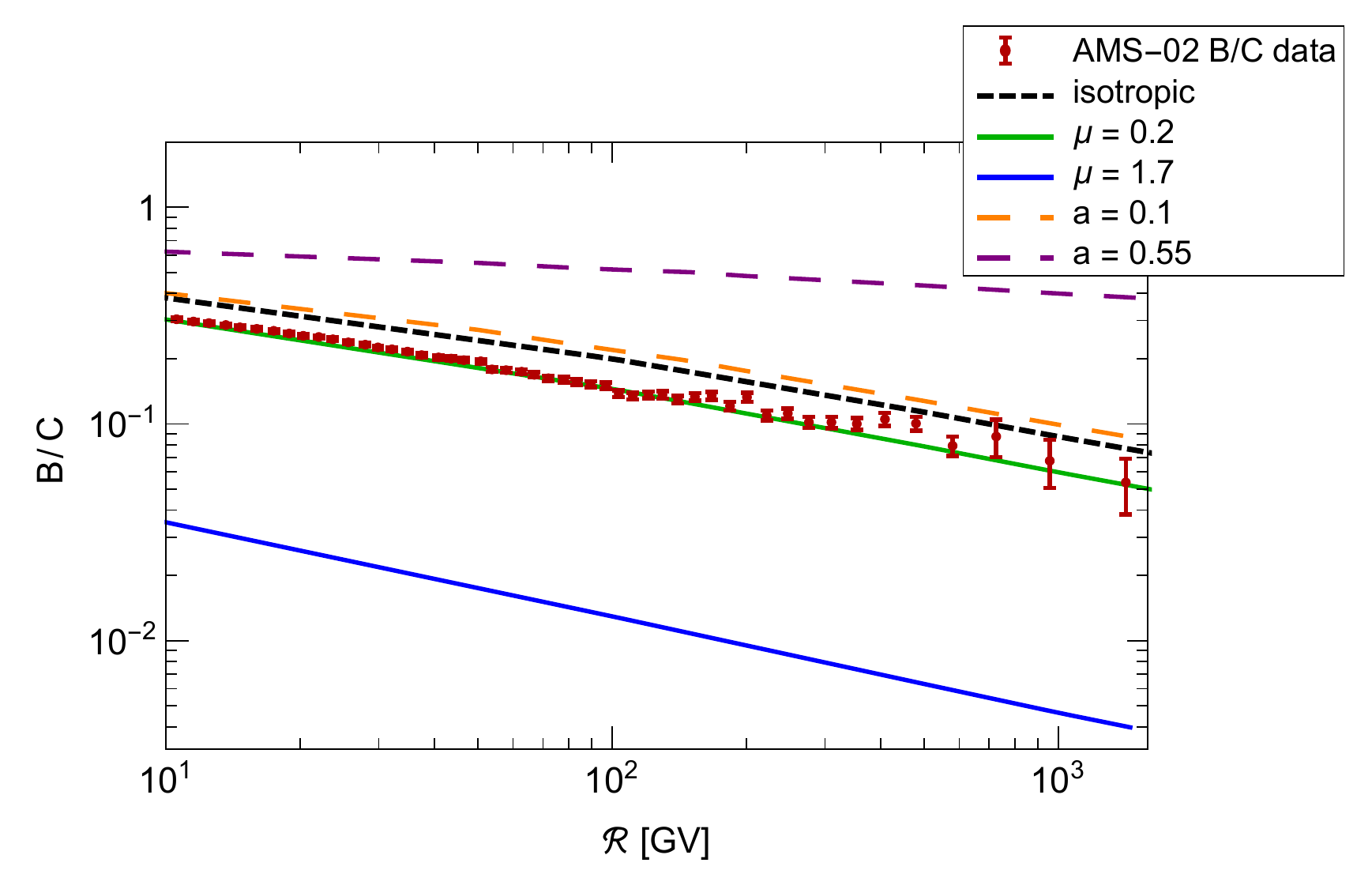}
	\caption{Left: The B/C spectrum as a function of the rigidity derived in the isotropic model and in different spatial dependent scenarios with the propagation parameters from table \ref{tab:bestBC}. Right: The B/C ratio derived with the propagation parameters given in table \ref{tab:propParameters} and $\delta=0.4$.}
	\label{fig:BCaniso}
\end{figure}

We want to study the effects of the spatial dependent diffusion model on the hadronic spectra.
Having neglected convection and reacceleration, the solution is applicable to the high energy part ($\geq \text{few tens}\ \GeVn$) of these spectra. 
We find that after propagation the flux is proportional to $\Phi\propto Q/K\propto \mathcal{R}^{-(\gamma_0+\delta)}$. This corresponds to the same result which is obtained in the isotropic model and satisfies our expectations from a phenomenological point of view as displayed in figure \ref{fig:BCaniso}. For the derivation of the B/C ratio we proceeded as described in detail in \cite{Kappl:2015bqa}. The primary spectra of the particle species responsible for boron production are determined from a fit to the data. The cross sections are taken from \cite{Webber03}.

Consequently, the B/C spectrum in the inhomogeneous diffusion model will be proportional to $\mathcal{R}^{-\delta}$. From previous studies we know that $\delta\approx 0.4$ is required to match the data \cite{Giesen:2015ufa,Evoli:2015vaa,Kappl:2015bqa}.
This is obviously in tension with the very small value of $\delta$ which is needed to describe the high energy part of the positron spectrum. However, the assumption that all kinds of cosmic rays propagate with the same parameters is under debate \cite{Johannesson:2016rlh,Cowsik:1975tj,Blasi:2009hv}. One possibility to justify the choice of different diffusion slopes for leptons and hadrons is provided by the Nested Leaky Box Model \cite{Cowsik:1975tj,Cowsik:2015yra}. 

The decreasing B/C ratio is explained using the discreteness of the sources and propagation features differing for the circumstellar and the interstellar medium. It has been studied \cite{2003LNP...598..171C} that in the vicinity of supernovae, serving as promising candidates for cosmic ray acceleration, the diffusion coefficient seems to increase with energy. Furthermore it has been conjectured that the energy dependence of the diffusion coefficient weakens for high energies, which is strongly supported by the observation of a spectral hardening in many cosmic ray spectra around rigidity $\mathcal{R}\sim \SI{400}{GV}$ \cite{Tomassetti:2012ga,Thoudam:2014sta,Bernard:2012pia}.

For the secondary production of positrons, nuclei of high energies ($\geq \SI{100}{\GeVn}$) are required as the positron carries away only a few percent of the energy of the parent nucleus. As those high energetic primary particles escape from the circumstellar regions quickly, the positron spectrum we observe is mainly produced in the interstellar medium. The harder spectrum of the primary particles is translated into the production spectrum of positrons which then propagate in the interstellar medium where the diffusion coefficient is energy independent.

In contrast, in the secondary production of boron nearly the full energy of the parent nucleus is adopted, so primaries and secondaries have approximately the same energy. The primary cosmic rays responsible for boron production are thus from a far lower energy regime than the ones producing positrons in the energy range of interest. As the escape time from the circumstellar regions decreases with energy $\tau\propto \mathcal{R}^{-\delta}$, the amount of produced secondaries also decreases with energy.
This gives rise to the observed decrease of the B/C ratio, while positrons do not adopt this spectral feature.

We find that the high energy parts of the spectra can be reproduced in principle in both the power-law and the exponentially inhomogeneous scenarios by choosing different values for \(\delta\) for leptons and hadrons. We expect that it is possible to fix the mismatch in the normalization for the B/C ratio by adjusting \(K_0\) appropriately. The remaining caveat is the bump in the low energy part of the positron spectrum. 
Our model is unable to describe this spectral feature.

\section{Conclusion}
\label{chap:conclusion}

The observation of cosmic rays offers the opportunity to study dark matter via indirect detection. 
A good signal to background ratio is expected for antiparticles. 
In this light we studied the origin of the observed positron excess by modifying the established two-zone diffusion model, giving rise to a significant change for the spectral features of the background prediction.

We presented an inhomogeneous propagation model with a diffusion coefficient that increases with galactic height. 
Our approach is well motivated from the spatial distribution of the magnetic field which again is responsible for the diffusion process. For the first time we derived an analytic solution for the inhomgeneous model. Considering two different cases we explored a diffusion coefficient depending on the galactic height first as a power-law and second as an exponential function. The modification reduces the propagation time scale such that energy losses are less relevant. In this model the high energy part of the measured positron spectrum can be reproduced with secondary positrons only. We also pointed out that this is a particular feature of the inhomogeneity. 
As a drawback, the model in such a setup is in contrast to the isotropic model no longer able to describe the low energy part of the observed positron spectrum.

As a conclusion we claim that the inhomogeneity strongly affects the spectral shape of leptonic spectra and that these effects have to be taken into account for constraining additional primary sources.

Our extended model can be tested experimentally. One feature of the inhomogeneous model is the prediction of a spectral softening in the positron spectrum. This offers a possibility to discriminate its spectral features from a pulsar contribution, where an exponential cutoff is expected. Furthermore the modification implies a reduction of the propagation time, such that the model can be constrained by analyzing the spectra of long lived radioactive particle species, e.g.\! $^{10}\text{Be}/^{9}\text{Be}$. Lastly an analysis of synchrotron radiation could quantify the spatial particle density distribution which provides another chance to discriminate between the different models.

In order to check consistency of the inhomogeneous model with other particle species we investigated an analytic derivation of the hadronic fluxes. We found that these spectra are not significantly affected by the modification. This can be understood as energy losses play a minor role for the propagation of nuclei. 
Further investigation is required to examine whether the high energy part of the positron spectrum and the B/C spectrum can be described with the same set of propagation parameters. However, in the literature the method of using the same propagation parameters for leptons and hadrons is under debate. If the model can further be extended to describe the positron spectrum at lower energies is beyond the scope of this work. 

Finally we want to stress how many exciting possibilities are currently offered by indirect detection for the exploration of dark matter. If we want to benefit from the precision data which is expected to be available within the next few years, it therefore becomes a major issue to understand the astrophysical backgrounds better. 
Going beyond the ordinary two-zone diffusion model a more realistic description of cosmic ray propagation has been achieved in our work. The resulting analytic expressions can easily be applied to future studies.

\section*{Acknowledgments}
This work has been supported by the German Science Foundation (DFG) within the SFB-Transregio TR33 ``The Dark Universe''. A. R. is partially supported by a fellowship of the Cusanuswerk. We thank V. Dogiel, D. Hooper and M. W. Winkler for helpful discussions and T. Kamae for providing the code to reproduce the results of \cite{Kamae:2006bf}.

\appendix
\section{Solving the Inhomogeneous Diffusion Equation for Positrons}
\label{app:solve}

We start from the simplified diffusion equation with energy losses and inhomogeneous diffusion
\begin{equation}
-\frac{\partial}{\partial z}\left[K(E,z)\frac{\partial}{\partial z}\mathcal{N}\right]+\frac{\partial}{\partial E}\left(b^\text{loss}\mathcal{N}\right)=Q(E,z).
\end{equation}
We define a coordinate transformation
\begin{equation}
\mathcal{F}=b^\text{loss}\mathcal{N}=-b_0E^2\mathcal{N},\qquad E\in [0,\infty]\mapsto \lambda=\frac{K_0}{b_0}\frac{E^{\delta-1}}{1-\delta}\in [\infty,0]
\end{equation}
and find with the source from \eqref{eq:sourcetermPos} a Fokker-Planck equation \cite{Baltz:1998xv}
\begin{equation}
\label{eq:diffeq2}
-\frac{\del}{\del z}\left[f(z)\frac{\del\mathcal{F}}{\del z}\right]+\frac{\del\mathcal{F}}{\del\lambda}=Q(\lambda,z)\, ,\qquad Q(\lambda,z)=2hq_0\frac{b_0}{K_0}\left(\frac{b_0}{K_0}(1-\delta)\lambda\right)^{\frac{2-\delta-\gamma_0}{\delta-1}}\delta(z-\xi).
\end{equation}
Choosing a general notation, $f(z)$ denotes the $z$-dependent factor of the diffusion coefficient \eqref{eq:DiffCoefficient} whose functional dependence is specified in the following sections. We have further introduced \(\xi\) for generality and will discuss it in more detail later.
The inhomogeneous partial differential equation \eqref{eq:diffeq2} can be solved by determining the Green's function $\mathcal{G}(\lambda,\lambda_s,z,z_s)$ which is defined by
\begin{equation}
-\frac{\del}{\del z}\left[f(z)\frac{\del \mathcal{G}}{\del z}\right]+\frac{\del \mathcal{G}}{\del\lambda}=\delta(\lambda-\lambda_s)\delta(z-z_s)\, .
\end{equation}
A Laplace transformation in \(\lambda\) results in an ordinary differential equation with $\mathcal{G}(\lambda,\lambda_s,z,z_s)\mapsto g(s,\lambda_s,z,z_s)$
\begin{equation}
\label{eq:greenLaplace4new}
-\frac{\del}{\del z}\left[f(z)\frac{\del g}{\del z}\right]+sg=\e^{-s\lambda_s}\delta(z-z_s)\, .
\end{equation}
On the next pages we present the solution for a power-law and an exponential dependence of $f(z)$.

\subsection{Power-Law Dependence}
\label{app:powerlaw}

Let
\begin{equation}
\label{eq:zdepDiffCoeff}
K(E,z)=K_0E^\delta \left|z\right|^\mu,\qquad 0\leq\mu < 2.
\end{equation}
The diffusion volume is restricted by $-L < z < L$. 
For symmetry reasons we can focus on $z>0$ in the derivation and extend the result to the full region in the end. 
The substitution $z=\chi^k$ with an arbitrary exponent $k$ allows to rewrite \eqref{eq:greenLaplace4new} with \eqref{eq:zdepDiffCoeff} to
\begin{equation}
\label{eq:greenLaplace3new}
\frac{1}{k^2}\chi^{k(\mu-2)}\left[\chi^2\del^2_\chi g+(1+k(\mu-1))\chi\del_\chi g\right]-sg=-e^{-s\lambda_s}\delta\left(\chi^k-\chi_s^k\right)\, .
\end{equation}
Here $\chi_s$ denotes $\chi(z=z_s)$ and at the boundary we define $\chi_L=\chi(z=L)$.
Equation \eqref{eq:greenLaplace3new} can be solved with a Fourier-Bessel expansion which respects the Dirichlet boundary conditions at \(z=L\)
\begin{equation}
\label{eq:ansatz}
g=\sum_{n=1}^\infty A_n \chi^l\mathcal{J}_m\left(\zeta_n\frac{\chi}{\chi_L}\right)\, .
\end{equation}
The coefficients $A_n$ depend on $s, \lambda_s$ and $\chi_s$, the second factor is an arbitrary power-law and the third one a Bessel function $\mathcal{J}_m$ of the first kind and $m^\text{th}$ order. The $\zeta_n$ denote the Bessel function's roots such that \eqref{eq:ansatz} fulfills the boundary condition $g(\chi_L) = 0$. Inserting \eqref{eq:ansatz} into \eqref{eq:greenLaplace3new} gives
\begin{equation}
\begin{split}
\label{eq:greenLaplace2}
&\sum_nA_n\bigg[\chi^2\mathcal{J}_m''+\overbrace{\left(2l+1+k(\mu-1)\right)}^{\stackrel{!}{=}1}\chi\mathcal{J}_m'+\bigg(l(l+k(\mu-1))-sk^2\overbrace{\chi^{-k(\mu-2)}}^{\stackrel{!}{=}\chi^2}\bigg)\mathcal{J}_m\bigg]\\
=&-e^{-s\lambda_s}|k|\chi^{-l-k(\mu-2)}\chi_s^{-k+1}\delta\left(\chi-\chi_s\right),
\end{split}
\end{equation}
with $\mathcal{J}_m'=\frac{\del}{\del\chi}\mathcal{J}_m\left(\zeta_n\frac{\chi}{\chi_L}\right)$.
For $k=\frac{2}{2-\mu}$ and $l=\frac{1-\mu}{2-\mu}$ the l.h.s.\! of \eqref{eq:greenLaplace2} is of the form of Bessel's differential equation\footnote{By definition a solution to Bessel's differential equation $x^2\frac{\dd^2 f}{\dd x^2}+x\frac{\dd f}{\dd x}+\left((cx)^2-m^2\right)f=0$ is given by a Bessel function of the first kind and of $m^\text{th}$ order $f=\mathcal{J}_m(cx)$. For a series representation see e.g. \cite{abramowitz}.}. The order of the Bessel function is identified to be $m=\pm l=\pm\frac{1-\mu}{2-\mu}$, where we have to choose the negative one, such that the solution decreases monotonically between zero and the first root.

Using Bessel's equation with these values one can get rid of the derivatives in \eqref{eq:greenLaplace2} 
\begin{equation}
\label{eq:greenLaplace11}
\sum_nA_n\left(-sk^2-\left(\frac{\zeta_n}{\chi_L}\right)^2\right)\chi^2\mathcal{J}_m\left(\zeta_n\frac{\chi}{\chi_L}\right)
=-e^{-s\lambda_s}|k|\chi^{2-l}\chi_s^{1-k}\delta\left(\chi-\chi_s\right).
\end{equation}
To find the expansion coefficients \(A_n\) we expand the r.h.s.\! of \eqref{eq:greenLaplace11} into a Fourier-Bessel series after dividing by $\chi^2$
\begin{equation}
\sum_n A_n\left(s+\Omega_n^2\right)\mathcal{J}_m\left(\zeta_n\frac{\chi}{\chi_L}\right)
=
\frac{\e^{-s\lambda_s}}{|k|}\chi_s^{1-k}\underbrace{\sum_j a_j\mathcal{J}_m\left(\zeta_j\frac{\chi}{\chi_L}\right)}_{\stackrel{!}{=}\chi^{-l}\delta(\chi-\chi_s)}\label{eq:greenLaplace6}\, .
\end{equation}
Here $\Omega_n=\frac{\zeta_n}{k\chi_L}$ was introduced as a shorthand notation.
The expansion coefficients $a_j$ can be derived respecting the symmetry of the setting to be 
\begin{equation}
\label{eq:an}
a_j=\frac{1}{\chi_L^2\left[\mathcal{J}_{m+1}\left(\zeta_j\right)\right]^2}\,\chi_s^{1-l}\mathcal{J}_m\left(\zeta_j\frac{\chi_s}{\chi_L}\right).
\end{equation}
As the Fourier-Bessel coefficients are linearly independent, the relation in \eqref{eq:greenLaplace6} holds for each order $n=j$ separately. Thus using \eqref{eq:an} we find an expression for each coefficient $A_n$
\begin{equation}
A_n=\frac{1}{|k|}\frac{\chi_s^{l}}{\chi_L^2}\frac{e^{-s\lambda_s}}{s+\Omega_n^2}\frac{\mathcal{J}_m\left(\zeta_n\frac{\chi_s}{\chi_L}\right)}{\left[\mathcal{J}_{m+1}\left(\zeta_n\right)\right]^2}\, .
\end{equation}
In total the solution to \eqref{eq:greenLaplace4new} is given by
\begin{equation}
\label{eq:linkHad}
g(s,\lambda_s,\chi,\chi_s)=\frac{1}{|k|}\frac{\left(\chi\chi_s\right)^{l}}{\chi_L^2}\sum_n\frac{e^{-s\lambda_s}}{s+\Omega_n^2}\frac{\mathcal{J}_m\left(\zeta_n\frac{\chi_s}{\chi_L}\right)\mathcal{J}_m\left(\zeta_n\frac{\chi}{\chi_L}\right)}{\left[\mathcal{J}_{m+1}\left(\zeta_n\right)\right]^2}\, .
\end{equation}
Taking the inverse Laplace transform \cite{oberhettinger2012tables} and resubstituting $\chi=\sqrt[k]{z}$ we find the Green's function
\begin{equation}
\label{eq:Greensfunction}
\mathcal{G}(\lambda,\lambda_s,z,z_s)
=\frac{\Theta(\lambda-\lambda_s)}{|k|}\frac{\sqrt[k]{zz_s}^{l}}{\sqrt[k]{L}^{2}}
\sum_n 
\frac{\mathcal{J}_m\left(\zeta_n\sqrt[k]{\frac{z_s}{L}}\right)\mathcal{J}_m\left(\zeta_n\sqrt[k]{\frac{z}{L}}\right)}{\left[\mathcal{J}_{m+1}\left(\zeta_n\right)\right]^2}e^{-\Omega_n^2(\lambda-\lambda_s)}
\end{equation}
where $\Theta(\lambda-\lambda_s)$ denotes the Heaviside step function. We can further simplify the spatial dependence by using the identities
\begin{equation}
\mathcal{J}_m\left(z\right)=\left(\frac{z}{2}\right)^m\frac{{}_0F_{1}\left(m+1;-\frac{1}{4}z^2\right)}{\Gamma(m+1)}\quad \forall\ m \notin -1,-2,\ldots,\qquad \Gamma(m+1)=m\Gamma(m)
\end{equation}
to obtain
\begin{equation}
\mathcal{G}(\lambda,\lambda_s,z,z_s)
=\frac{2\Theta(\lambda-\lambda_s)}{|2-\mu|}
\sum_n 
\frac{{}_0F_1\left(\frac{1}{2-\mu};-\frac{1}{4}\zeta_n^2\left(\frac{z_s}{L}\right)^{2-\mu}\right){}_0F_1\left(\frac{1}{2-\mu};-\frac{1}{4}\zeta_n^2\left(\frac{z}{L}\right)^{2-\mu}\right)}
{L\zeta_n^2\ {}_0F_1\left(1+\frac{1}{2-\mu};-\frac{1}{4}\zeta_n^2\right)^2}
e^{-\Omega_n^2(\lambda-\lambda_s)}\, .
\end{equation}

The inhomogeneous solution $\mathcal{F}$ which is related to the particle density $\mathcal{N}(E,z)$ can be found performing a convolution of the Green's function with the inhomogeneous source term
\begin{align}
\label{eq:green2}
\mathcal{F}&=\intop_{\infty}^0\dd \lambda_s\intop_{-\infty}^{\infty}\dd z_s\,\mathcal{G}(\lambda,\lambda_s,z,z_s)Q(\lambda_s,z_s)\\
&=\frac{2}{\left|2-\mu\right|}\frac{2h}{L}q_0
\sum_n 
\frac{{}_0F_1\left(\frac{1}{2-\mu};-\frac{1}{4}\zeta_n^2\left(\frac{\xi}{L}\right)^{2-\mu}\right){}_0F_1\left(\frac{1}{2-\mu};-\frac{1}{4}\zeta_n^2\left(\frac{z}{L}\right)^{2-\mu}\right)}
{\zeta_n^2\ {}_0F_1\left(1+\frac{1}{2-\mu};-\frac{1}{4}\zeta_n^2\right)^2}
\nonumber\\
&\phantom{=}\cdot \intop_{\lambda}^0\dd \lambda_s
\frac{b_0}{K_0}\left(\frac{b_0}{K_0}(1-\delta)\lambda_s\right)^{\frac{2-\delta-\gamma_0}{\delta-1}}
\e^{-\Omega_n^2(\lambda-\lambda_s)}\\
&=\frac{2}{\left|2-\mu\right|}\frac{2h}{L}q_0
\sum_n 
\frac{{}_0F_1\left(\frac{1}{2-\mu};-\frac{1}{4}\zeta_n^2\left(\frac{\xi}{L}\right)^{2-\mu}\right){}_0F_1\left(\frac{1}{2-\mu};-\frac{1}{4}\zeta_n^2\left(\frac{z}{L}\right)^{2-\mu}\right)}
{\zeta_n^2\ {}_0F_1\left(1+\frac{1}{2-\mu};-\frac{1}{4}\zeta_n^2\right)^2}
\nonumber\\
&\phantom{=}\cdot \frac{E^{1-\gamma_0}}{1-\gamma_0}\ {}_1F_1\left(1,\frac{\gamma_0-\delta}{1-\delta};
-\Omega_n^2\frac{K_0}{b_0}\frac{E^{\delta-1}}{1-\delta}\right)
\end{align}
where we resubstituted \(\lambda\) to \(E\) after solving the integral in the last step.

The number density taking the symmetry in \(z\) into account is finally given by
\begin{equation}
\begin{split}
\mathcal{N}(E,z)&=\frac{2}{\left|2-\mu\right|}\frac{2h}{L}\frac{q_0}{b_0}
\sum_{n=1}^{\infty} 
\frac{{}_0F_1\left(\frac{1}{2-\mu};-\frac{1}{4}\zeta_n^2\left(\frac{\xi}{L}\right)^{2-\mu}\right){}_0F_1\left(\frac{1}{2-\mu};-\frac{1}{4}\zeta_n^2\left(\frac{z}{L}\right)^{2-\mu}\right)}
{\zeta_n^2\ {}_0F_1\left(1+\frac{1}{2-\mu};-\frac{1}{4}\zeta_n^2\right)^2}\\
&\phantom{=}\cdot \frac{E^{-(\gamma_0+1)}}{\gamma_0-1}\ {}_1F_1\left(1,\frac{\gamma_0-\delta}{1-\delta};
-\Omega_n^2\frac{K_0}{b_0}\frac{E^{\delta-1}}{1-\delta}\right)
\, ,
\end{split}
\end{equation}
with $\Omega_n=\frac{2-\mu}{2}\zeta_n L^{-\frac{2-\mu}{2}}$. We can obtain the flux at earth by setting \(z=\xi=0\). This is in principle possible but will lead to convergence problems for the sum, as \(\zeta_n\mapsto\infty\) for \(n\mapsto \infty\) and \({}_0F_1\left(\alpha;0\right)=1\) but \({}_0F_1\left(\alpha;-x^2\right)\mapsto 0\) for \(x\mapsto \infty\). We therefore set \(z=\xi\neq 0\) for some small value \(\xi\) which acts as a cutoff and ensures further that the diffusion coefficient \(K(E,z)\) is non-vanishing in the galactic disk. The result depends on the concrete choice for \(\xi\) which can be understood as parameter for the diffusion strength in the galactic plane.

We finally find
\begin{equation}
\label{eq:phiAni}
\begin{split}
\Phi(E)&=\frac{\beta}{4\pi}\frac{2}{\left|2-\mu\right|}\frac{2h}{L}\frac{q_0}{b_0}\frac{E^{-(\gamma_0+1)}}{\gamma_0-1}
\sum_{n=1}^{\infty} 
\frac{{}_0F_1\left(\frac{1}{2-\mu};-\frac{1}{4}\zeta_n^2\left(\frac{\xi}{L}\right)^{2-\mu}\right)^2}
{\zeta_n^2\ {}_0F_1\left(1+\frac{1}{2-\mu};-\frac{1}{4}\zeta_n^2\right)^2}\\
&\phantom{=}\cdot 
{}_1F_1\left(1,\frac{\gamma_0-\delta}{1-\delta};
-\Omega_n^2\frac{K_0}{b_0}\frac{E^{\delta-1}}{1-\delta}\right)
\, .
\end{split}
\end{equation}
It is also possible to use the source term from \eqref{eq:sourceDisk} where the sources are distributed homogeneously allover the galactic disk and in that case
\begin{equation}
\label{eq:foriso}
\begin{split}
\mathcal{N}(E,z)&=\frac{2}{\left|2-\mu\right|}\frac{2h}{L}\frac{q_0}{b_0}
\sum_{n=1}^{\infty} 
\frac{{}_0F_1\left(1+\frac{1}{2-\mu};-\frac{1}{4}\zeta_n^2\left(\frac{h}{L}\right)^{2-\mu}\right){}_0F_1\left(\frac{1}{2-\mu};-\frac{1}{4}\zeta_n^2\left(\frac{z}{L}\right)^{2-\mu}\right)}
{\zeta_n^2\ {}_0F_1\left(1+\frac{1}{2-\mu};-\frac{1}{4}\zeta_n^2\right)^2}\\
&\phantom{=}\cdot \frac{E^{-(\gamma_0+1)}}{\gamma_0-1}\ {}_1F_1\left(1,\frac{\gamma_0-\delta}{1-\delta};
-\Omega_n^2\frac{K_0}{b_0}\frac{E^{\delta-1}}{1-\delta}\right)
\, ,
\end{split}
\end{equation}
with the help of
\begin{equation}
2\intop_0^h \dd z_s\ {}_0F_1\left(\frac{1}{2-\mu};-\frac{1}{4}\zeta_n^2\left(\frac{z_s}{L}\right)^{2-\mu}\right)
=2h\ {}_0F_1\left(1+\frac{1}{2-\mu};-\frac{1}{4}\zeta_n^2\left(\frac{h}{L}\right)^{2-\mu}\right)\, .
\end{equation}

\subsubsection{Reduction to Isotropic Case}
\label{app:isotropic}

We can directly set the inhomogeneity to zero in \eqref{eq:foriso}. We find with \cite{abramowitz}
\begin{equation}
{}_0F_1\left(\frac{1}{2},-\frac{1}{4}z^2\right)=\cos(z),\qquad 
{}_0F_1\left(\frac{3}{2},-\frac{1}{4}z^2\right)=\frac{\sin(z)}{z}
\end{equation}
and the fact that \(\Omega_n=k_n\) for \(\mu=0\)
\begin{equation}
\begin{split}
\mathcal{N}(E,z)&=\frac{2q_0}{b_0}\frac{E^{-(\gamma_0+1)}}{\gamma_0-1}\sum_{n=0}^\infty\frac{\cos\left(\pi\left(n+\frac{1}{2}\right)\frac{z}{L}\right)\sin\left(\pi\left(n+\frac{1}{2}\right)\frac{h}{L}\right)}{\pi(n+\frac{1}{2})}\\
&\phantom{=}\cdot {}_1F_1\left(1,\frac{\gamma_0-\delta}{1-\delta};-\frac{\pi^2\left(n+\frac{1}{2}\right)^2}{L^2}\frac{K_0}{b_0}\frac{E^{\delta-1}}{1-\delta}\right)\, .
\end{split}
\end{equation}
This is exactly the known result from \cite{1974Ap&SS..29..305B} and a nice consistency check of our approach.

\subsection{Exponential Dependence}
\label{chap:Zexp}

Motivated by the spatial structure of the galactic magnetic field we investigate an exponentially increasing diffusion coefficient of the form
\begin{equation}
\label{eq:expDiffCoeff}
K(E,z)=K_0E^\delta e^{az}\, .
\end{equation}
We need to find the Green's function for
\begin{equation}
\label{eq:diffusionexpz}
-\frac{\del}{\del z}\left[e^{az}\frac{\del}{\del z}\mathcal{G}\right]+\frac{\del\mathcal{G}}{\del\lambda}=\delta(z-z_s)\delta(\lambda-\lambda_s).
\end{equation}
A Laplace transformation simplifies the problem to an ordinary differential equation for the transformed Green's function $g(s,\lambda_s,z,z_s)$. Furthermore a general substitution $e^{z}=\chi^k$ yields
\begin{equation}
\label{eq:expZ1}
\chi^{ak}\left[\chi^2\del_\chi^2 g+(ak+1)\chi\del_\chi g\right]-k^2sg=-e^{-s\lambda_s}|k|\chi_s\delta\left(\chi-\chi_s\right)
\end{equation}
with the same notation as in \ref{app:powerlaw}. The ansatz \eqref{eq:ansatz} allows to transform the differential equation \eqref{eq:expZ1} such that is resembles Bessel's differential equation
\begin{equation}
\begin{split}
&\sum_n A_n\left[\chi^2\mathcal{J}_m''
+(2l+ak+1)\chi\mathcal{J}_m'
+\left[-sk^2\chi^{-ak}+l(l+ak)\right]\mathcal{J}_m
\right]\\
=&-e^{-s\lambda_s}|k|\chi_s\chi^{-(l+ak)}\delta\left(\chi-\chi_s\right).\label{eq:expZ2}
\end{split}
\end{equation}
For $k=-\frac{2}{a}$, $l=1$ and $m=\pm l = \pm 1$ we can use Bessel's differential equation to get rid of the terms containing derivatives. Once more we expand the r.h.s.\! of \eqref{eq:expZ2} into a Fourier-Bessel series and determine the coefficients $A_n$ for each term separately. For the Laplace transformed Green's function we find
\begin{equation}
g(s,\lambda_s,\chi,\chi_s)=\frac{2}{|k|}\frac{\chi_s\chi}{\chi_L^2}\sum_n
\frac{e^{-s\lambda_s}}{s+\Lambda_n^2}\frac{\mathcal{J}_1\left(\zeta_n\frac{\chi_s}{\chi_L}\right)\mathcal{J}_1\left(\zeta_n\frac{\chi}{\chi_L}\right)}{\left[\mathcal{J}_2(\zeta_n)\right]^2}
\end{equation}
with $\Lambda_n=-\frac{\zeta_n}{k\chi_L}=\frac{a}{2}\zeta_n e^{\frac{a}{2}L}$. Taking the inverse Laplace transform and resubstituting $\chi=e^{-\frac{a}{2}z}$ yields
\begin{equation}
\mathcal{G}(\lambda,\lambda_s,z,z_s)
=\Theta(\lambda-\lambda_s)2\vert a\vert e^{a\left(L-\frac{z+z_s}{2}\right)}
\sum_n
\frac{\mathcal{J}_1\left(\zeta_ne^{\frac{a}{2}(L-z_s)}\right)\mathcal{J}_1\left(\zeta_ne^{\frac{a}{2}(L-z)}\right)}{\left[\mathcal{J}_2(\zeta_n)\right]^2}e^{-\Lambda_n^2(\lambda-\lambda_s)}\ .\label{eq:greenExp}
\end{equation}
Including the source term, the solution for the number density is derived similarly as in \eqref{eq:green2} giving
\begin{equation}
\begin{split}
\mathcal{N}(E,z)&=2h\frac{q_0}{b_0}|a|e^{a\left(L-\frac{z+\xi}{2}\right)}
\sum_{n=1}^\infty\frac{\mathcal{J}_1\left(\zeta_ne^{\frac{a}{2}(L-\xi)}\right)\mathcal{J}_1\left(\zeta_ne^{\frac{a}{2}(L-z)}\right)}{\left[\mathcal{J}_2(\zeta_n)\right]^2}\\
&\phantom{=}\cdot \frac{E^{-(\gamma_0+1)}}{\gamma_0-1}{}_1F_1\left(1,\frac{\gamma_0-\delta}{1-\delta};-\Lambda_n^2\frac{K_0}{b_0}\frac{E^{\delta-1}}{1-\delta}\right).
\end{split}
\end{equation}
We can further set \(z=\xi=0\) and obtain
\begin{equation}
\Phi(E)=\frac{\beta}{4\pi}2h\frac{q_0}{b_0}|a|e^{aL}
\frac{E^{-(\gamma_0+1)}}{\gamma_0-1}
\sum_{n=1}^\infty\left[
\frac{\mathcal{J}_1\left(\zeta_ne^{\frac{a}{2}L}\right)}
{\mathcal{J}_2(\zeta_n)}
\right]^2
{}_1F_1\left(1,\frac{\gamma_0-\delta}{1-\delta};-\Lambda_n^2\frac{K_0}{b_0}\frac{E^{\delta-1}}{1-\delta}\right)\, .
\end{equation}
\bibliography{thesis_refs}
\bibliographystyle{ArXiv}
\end{document}